\documentclass{aa}
\usepackage{natbib,graphicx}
\usepackage{subfig}
\usepackage{textcomp}
\usepackage{rotating}
\usepackage{amssymb,amsmath}
\usepackage{txfonts}
\begin{document}

\title{Early-type galaxies with core collapse supernovae}

\author{A. A. Hakobyan\inst{1}
\and  A. R. Petrosian\inst{1}
\and B. McLean\inst{2}
\and D. Kunth\inst{3}
\and R. J. Allen\inst{2}
\and M. Turatto\inst{4}
\and R. Barbon\inst{5}
}
\institute{
Byurakan Astrophysical Observatory and Isaac Newton Institute of Chile, Armenian Branch, Byurakan 378433, Armenia\\
\email{hakartur@rambler.ru}
\and
Space Telescope Science Institute, 3700 San Martin Drive, Baltimore, MD 21218, USA
\and
Institut d'Astrophysique de Paris (UMR 7095: CNRS \& Universit\'e Pierre
et Marie Curie), 98bis Bd Arago, F--75014 Paris, France
\and
INAF, Osservatorio Astrofisico  di Catania via Santa Sofia 78, 95123 Catania, Italy
\and
Dipartimento di Astronomia, Universita' di Padova, vicolo dell'Osservatorio 2, 35122, Padova, Italy
}

\date{Received 20 March 2008 / Accepted 16 June 2008}

\abstract
{}
{It is widely accepted that the progenitors of core collapse SNe are young massive
stars and therefore their host galaxies are mostly spiral or irregular galaxies
dominated by a young stellar population. Surprisingly, among morphologically classified
hosts of core collapse SNe, we find 22 cases where the host has been classified
as an Elliptical or S0 galaxy.}
{To clarify this apparent contradiction, we carry out a detailed morphological
study and an extensive literature search for additional information on the
sample objects.}
{Our results are as follows:

1.	Of 22 ``early type'' objects, 17 are in fact misclassified spiral galaxies,
one is a misclassified irregular, and one is a misclassified ring galaxy.

2. Of the 3 objects maintaining the early type classification, one (NGC2768) is a suspected merger remnant,
another (NGC4589) is definitely a merger, and the third (NGC2274) is in close interaction.
The presence of some amount of young stellar population in these galaxies is therefore not unexpected.}
{These results confirm the presence of a limited, but significant, number of core collapse SNe
in galaxies generally classified of early type. In all cases, anyway, there are independent indicators
of the presence in host galaxies of recent star formation due to merging or gravitational interaction.}


\maketitle

\section{Introduction}

As their cosmological importance as distance indicators continues to grow,
a detailed understanding of the origin of supernovae (hereafter SNe) is becoming one of the
more perplexing problems in modern astronomy. Especially in the last two decades,
improvements in observational capabilities and the broader spectral range of
SNe studies have resulted in an unexpected diversity of the observed
properties of SNe leading to the introduction of new types and subtypes.
These are likely to be related to different physical parameters of the stellar
progenitors of SNe (possibly leading to observable differences in the details of
their explosion mechanisms), and to differences in the properties of their immediate
environment (e.g., \citealp{TBP07}). Current limits on our
understanding of these issues make it difficult to come to firm conclusions on the
role of SNe in several fields like the chemical evolution of galaxies, the production
of the stellar remnants and the origin of cosmic rays. Perhaps most vexing
are the doubts that remain on the reliability  with which SNe can be used as
distance indicators up to cosmological distances.

It is widely accepted that SNe result from two major explosion mechanisms:
core collapse in massive stars ($M~\geq~8M_\odot$) and thermonuclear runaway of in a WD
approaching the Chandrasekar limit by accretion from a companion. The focus of this
paper is on SNe formed via core collapse (hereafter CCSNe). Different configurations
of the progenitor at the moment of explosion, different energetics associated with
the event, and possible interaction of the ejecta with circumstellar material produce
a large variety of characteristics, presently classified as SNe of types Ib and Ic
and several sub-types of type~II. Spectacular evidence for the core collapse nature
of SNe II comes from the detection of neutrinos of SN1987A in LMC
but indication that CCSNe have massive progenitors also comes from about a dozen
cases of direct identification of core collapse SN progenitors (e.g., \citealp{vD05,HSC06}),
from theoretical modeling (e.g., \citealp{WW95,HWF03}), and  from studies of the
environment of these SNe.
Their association with spiral arms in galaxies (e.g., \citealp{MvdB76,BTF94}),
with massive star-forming regions (e.g., \citealp{vD92}), and with local young
stellar population (e.g., \citealp{vDPB99}) is well known. This is consistent
with the dependence of the rate of explosions with the morphological types of the
host galaxies. Many studies have concluded that types II and Ib/c SNe occur only
in spiral and irregular galaxies, and mostly in those of the latest types (e.g., \citealp{CET99}).
In other words different studies show  that the progenitor stars of CCSNe are
young massive stars born in recent episodes of star formation in their host galaxies
(e.g., \citealp{H03}), with possible differences in their nature (e.g., \citealp{vD+03}).
Conversely, the apparent certainty of the association between CCSNe and young massive stars has
led to use CCSNe as tracers of recent star formation in galaxies
(e.g., \citealp{Nav+01,Petr+05}).

It is therefore puzzling that several authors have reported the presence of CCSNe
in E or S0 galaxies. In three articles,
\citeauthor*{vdBLF02} (\citeyear{vdBLF02,vdBLF03,vdBLF05}, hereafter vdBLF)
discussed the morphologies of host galaxies in the Lick Observatory Supernova Search
(LOSS; e.g., \citealp{FLT+01}) and report 8 cases of E and S0 galaxies harboring
CCSNe. In a new examination of the objects listed in the latest version
(October 30, 2007) of the Asiago Supernova catalog (ASC; e.g., \citealp{BBCT99})
we discovered 14 more cases of CCSNe host galaxies classified as type E or S0.
If demonstrated, the presence of CCSNe in early type galaxies should be associated
with the population of the \textit{prompt} SN~Ia advocated by \citealp{MDVP06} to explain
the excess of SN~Ia events observed in radio-loud ellipticals.
This excess is attributed to recent episodes of massive star formations which are
expected to produce CCSNe although in still debated number (\citealp{GRD08}).

It is known that type Ib/c SNe can be misclassified as SN~Ia because of the similarity
of their spectra at some epoch. A recent example is SN2004aw (\citealp{TPMV+06}).
It is not excluded therefore that some less extensively studied SN~Ia reported in early type galaxies,
might indeed be misclassified SN~Ib/c. However, since to our knowledge
a statistics of these cases is not available, we neglect such possible effect in what follows.
In this paper we study in detail all 22 cases with the aim of understanding the
reliability of the host galaxy classification. The sample is presented in more
detail in Sect.~2. The results of our study of individual host galaxies are
presented in Sect.~3, and in Sect.~4 we discuss and summarize our results.
Throughout this article we have assumed a value of
$H_0=75 \,\rm km \,s^{-1} \,Mpc^{-1}$ for the Hubble constant.

\section{The sample}

The three papers of vdBLF discussed the morphological classification of more than 600
host galaxies of SNe. This classification was done on the DDO system using images from
the 0.75~cm Katzman Automatic Imaging Telescope during the course of the
Lick Observatory Supernova Search as well as from the POSS-I, POSS-II, or SERC-J
and SERC-R blue and red exposures.
The Asiago Supernova Catalog\footnote{http://web.oapd.inaf.it/supern/cat.} referenced above
has presently (October 30, 2007) accumulated data for about 4300 SNe and their
host galaxies and it includes (among other parameters) the morphological classes of the hosts.
These come from various sources, mostly from RC3 (\citealp{dVdVC+91})
and LEDA (70\% and 20\% respectively).
It is interesting to note that the classifications are not always in agreement:
of the 8 cases reported by vdBLF four have been classified as spirals
in the ASC; conversely, three galaxies of the 14 from the ASC are classified as spirals
(or possible spirals) by vdBLF. We have therefore considered the full list of 22 cases
for our study for which an Elliptical or S0 type parent galaxy is suspected.
It is worth mentioning that we have not considered S0/a type galaxies for this study.
There are two crucial reasons for this decision: firstly S0/a type galaxies are well known
to have an increasingly large young stellar population (e.g., \citealp{NvdHS+05}) and can harbor
CCSN progenitors; secondly the distinction between S0 and S0/a's is subtle and
give rise to possible misclassification (e.g., \citealp{NLBC+95}). It is interesting to note that vdBLF
did not report any CCSN in S0/a type galaxy but in ASC there are 15 such galaxies with 15 CCSNe.
Table~\ref{tabl1}
presents the list of 22 cases with CCSNe identified in E or S0 type galaxies.
The first column gives the name of the
host galaxy sorted according to R.A., the second and third columns list the supernovae
and their classifications from ASC. In three cases two SNe events were discovered in
the same galaxy and it is remarkable that in one (NGC2836) both events where
core collapse explosions. Columns~4 and 5 give the morphological types of the hosts
according to vdBLF and to the ASC where ``:'' denote uncertain classifications and ``?''
very uncertain ones. Our remarks and the new galaxy types are given in column~6.

\begin{table*}[t]
  \caption{The list of early type galaxies with core collapse supernovae.}
  \label{tabl1}
  \centering
  \begin{tabular}{l l l l l l}
  \hline\hline
  \multicolumn{1}{c}{Galaxy} & \multicolumn{1}{c}{SN} & \multicolumn{1}{c}{SN type} & \multicolumn{2}{c}{Morphology} & \multicolumn{1}{c}{New galaxy types and Remarks} \\
  & & & \multicolumn{1}{c}{vdBLF} & \multicolumn{1}{c}{ASC} & \\
  \hline
   \multicolumn{1}{c}{(1)} & \multicolumn{1}{c}{(2)} & \multicolumn{1}{c}{(3)} & \multicolumn{1}{c}{(4)} & \multicolumn{1}{c}{(5)} & \multicolumn{1}{c}{(6)} \\
  \hline
   IC1529 & 2006du & II &  & S0 & Sbc type galaxy with LINER type AGN. \\
    & & & & & HI, 1.4~GHz radio source. \\
   NGC774 & 2006ee & II &  & S0 & S0. Embryonic spiral arm exist. Residual star formation. \\
   NGC838 & 2005H & II &  & S0: pec & Mrk1022. Im type galaxy. HI, 1.4~GHz radio source. \\
   MCG-01-07-35 & 2002aq & II & SBab & SB0 pec & Barred-Ring galaxy. HI source. \\
   PGC10652 & 2006ab & Ic &  & E? & Sb type galaxy. HI source. \\
   IC1861 & 1999eg & II &  & S0 & Sbc type galaxy. HI source. \\
   NGC1260 & 2006gy & IIn? &  & S0: & Sa. Extended $\rm H\alpha$ emission also at the SNe position. \\
   IC340 & 2002jj & Ic & S pec? & S0: & Sc type galaxy. HI source. \\
   UGC2836 & 2001I & IIn & E1 & S0 & Mrk1405. Sa type galaxy. HI, $\rm H_{2}$, CO and \\
    & 2003ih & Ib/c &  &  & 1.4~GHz radio source. \\
   UGC3432 & 2003kb & Ic: & S0/Sb & Scd: & Sc type galaxy. HI source. \\
    & 1996bv & Ia &  &  &  \\
   NGC2274 & 2005md & II: &  & E & E type galaxy in interaction. HI source. \\
   NGC2768 & 2000ds & Ib & E3/Sa & E6 & S0 type galaxy with LINER type AGN. HI, CO and \\
    & & & & & 1.4~GHz radio source. A possible merger remnant. \\
   IC2461 & 2002bx & II & S0 & Sb & Sbc type galaxy. HI and 1.4~GHz radio source \\
    & & & & & with emission line spectrum. \\
   VIIIZw140 & 2004X & II & E3 &  & Spiral galaxy? HI and 1.4~GHz radio source. \\
   NGC3720 & 2002at & II: & E1 & Sa: & Sa type galaxy. HI and 1.4~GHz radio source \\
    & & & & & with emission line spectrum. \\
   IC3203 & 2003ac & IIb: & S0 & Sb & Sb type galaxy. HI source. \\
   ESO506-G11 & 2005an & II: &  & S0 & Sb type galaxy. \\
   NGC4589 & 2005cz & Ib &  & E2 & Merger with LINER type AGN. HI and \\
    & & & & & 1.4~GHz radio source. \\
   ESO576-G40 & 2003am & II & S0 t &  & SBc type galaxy. HI source with \\
    & 1997br & Ia pec &  &  & emission line spectrum. \\
   MAPS-NGP O\_442\_2131626 & 2004V & II & E0 &  & Sa type galaxy. \\
   UGC12044 & 2002hz & Ib/c & Sb/S0 & Sab & Sab type galaxy. HI source. \\
   MCG-02-57-22 & 2001hh & II & Sa & SB0: & Sa type galaxy. \\
  \hline
  \end{tabular}
\end{table*}

\begin{figure*}[t]
\centering
\subfloat{
\label{sub1.1} 
\includegraphics[angle=90, width=0.5\hsize]{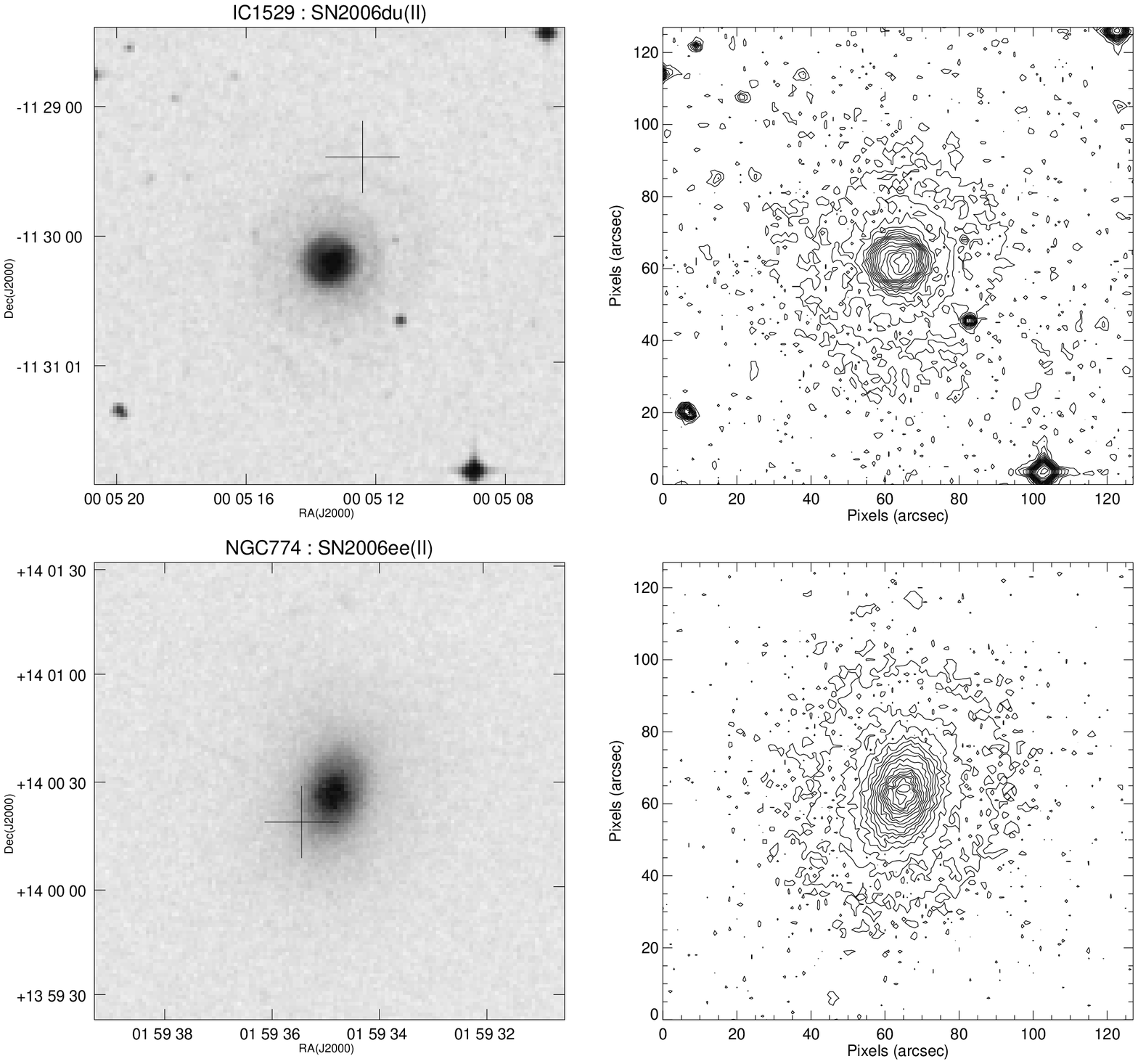}} \\
\subfloat{
\label{sub1.2} 
\includegraphics[angle=90, width=0.5\hsize]{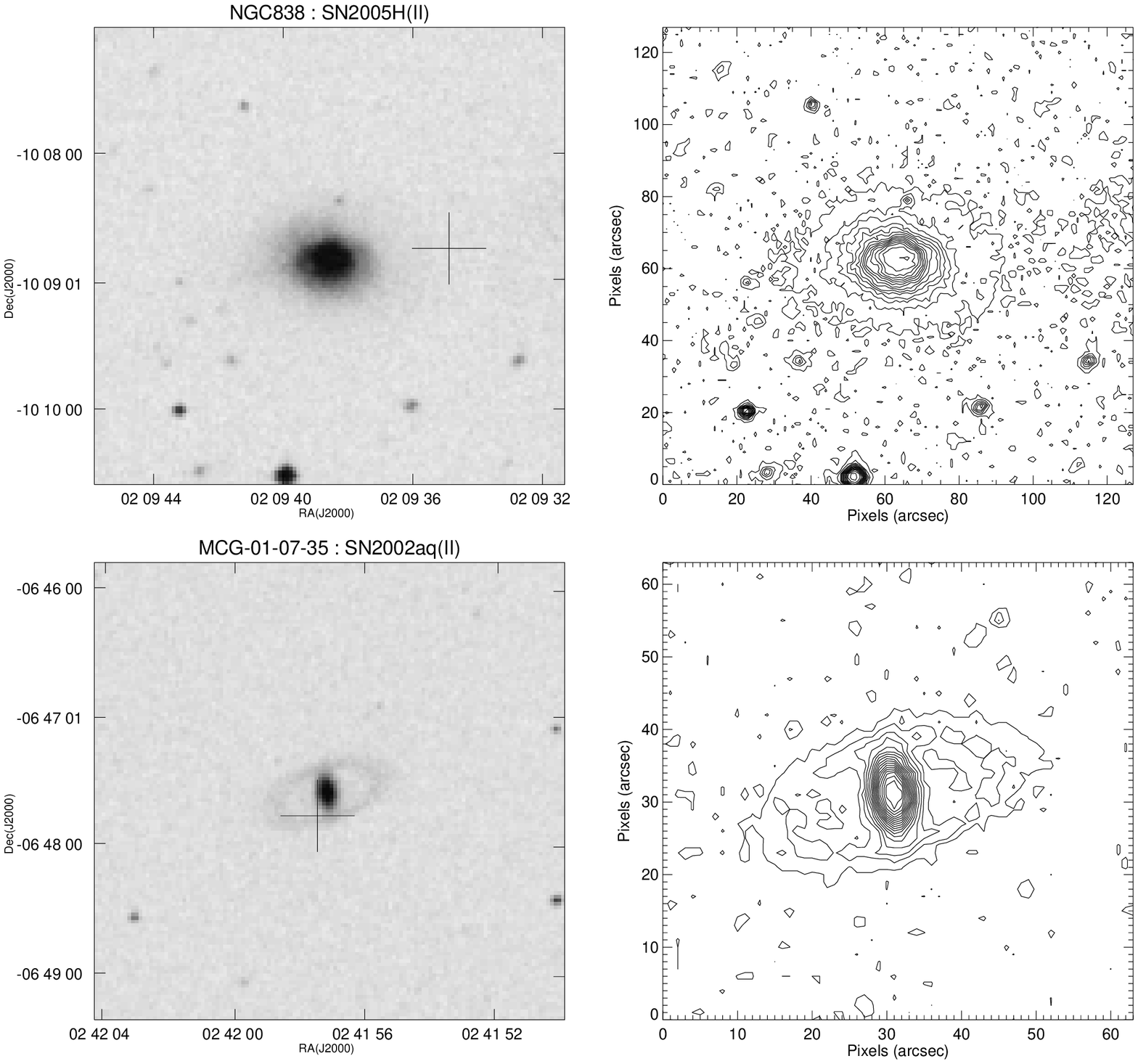}}
\caption{(a) Gray-scale and isophotal maps of early type galaxies with core collapse
supernovae. The positions of SNe are marked by ``+'' cross sign.}
\label{sn1} 
\end{figure*}
\begin{figure*}[t]
\ContinuedFloat
\centering
\subfloat{
\label{sub2.1} 
\includegraphics[angle=90, width=0.5\hsize]{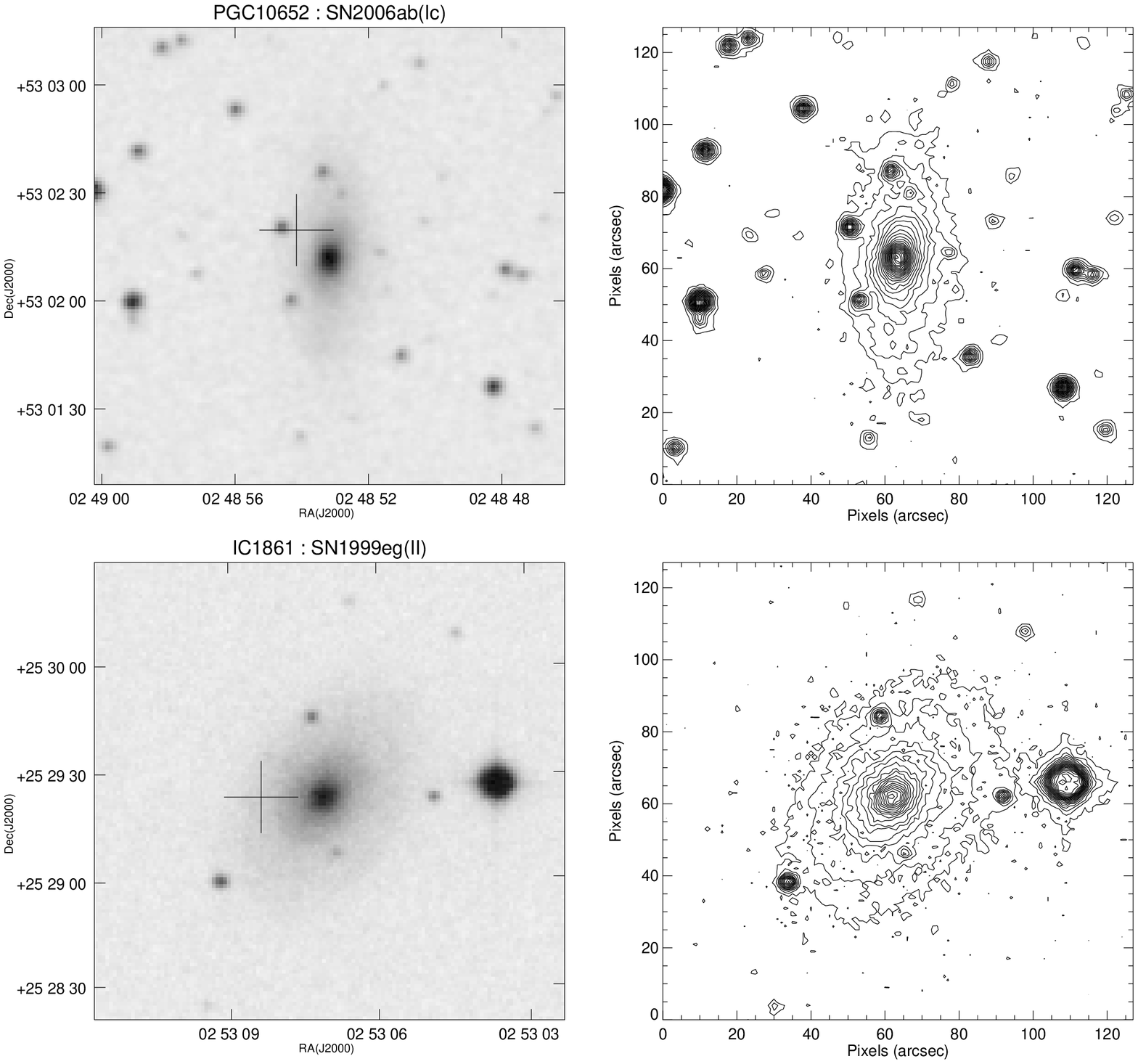}} \\
\subfloat{
\label{sub2.2} 
\includegraphics[angle=90, width=0.5\hsize]{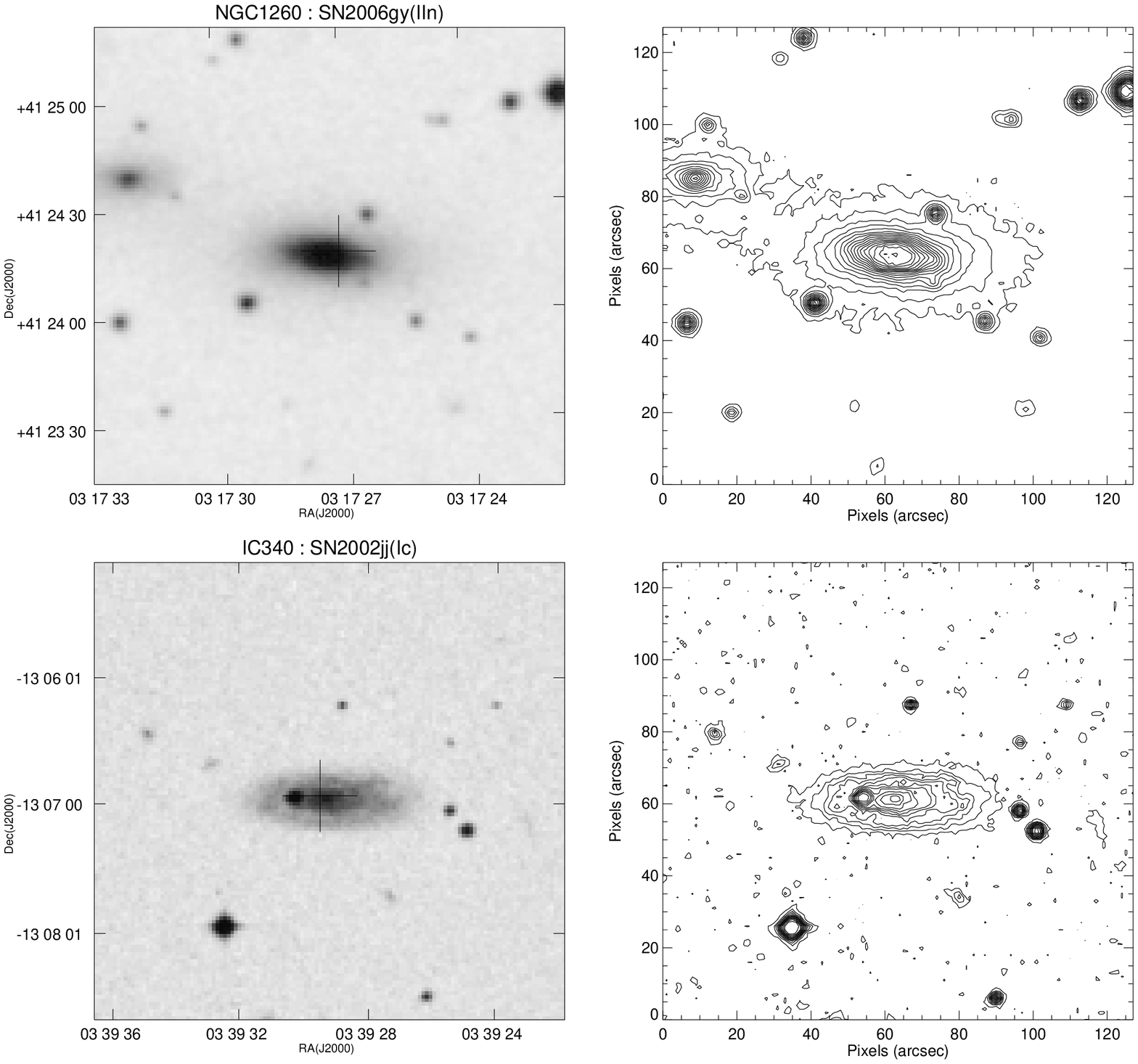}}
\caption{(b) Continued.}
\label{sn2} 
\end{figure*}
\begin{figure*}[t]
\ContinuedFloat
\centering
\subfloat{
\label{sub3.1}
\includegraphics[angle=90, width=0.5\hsize]{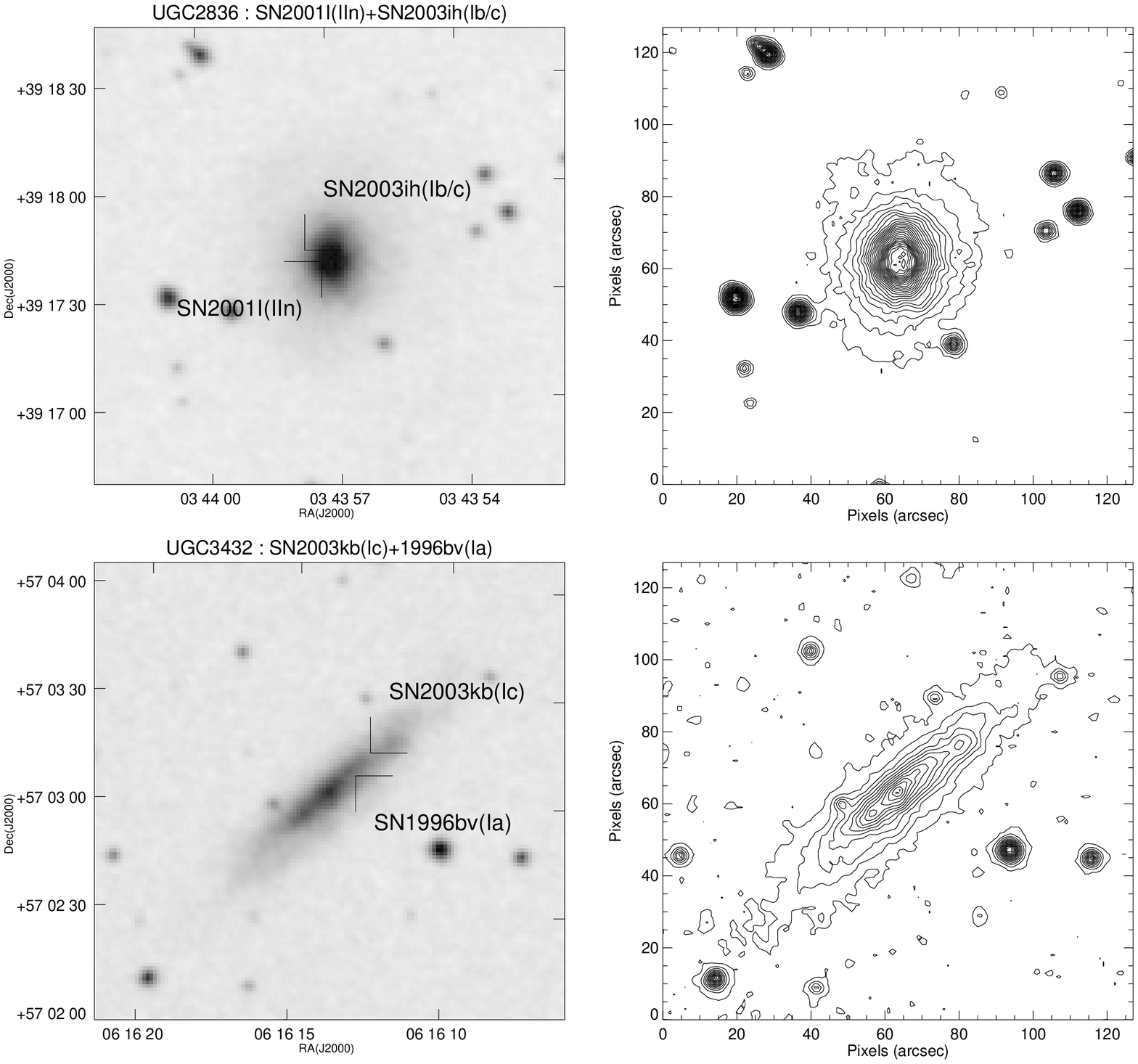}} \\
\subfloat{
\label{sub3.2}
\includegraphics[angle=90, width=0.5\hsize]{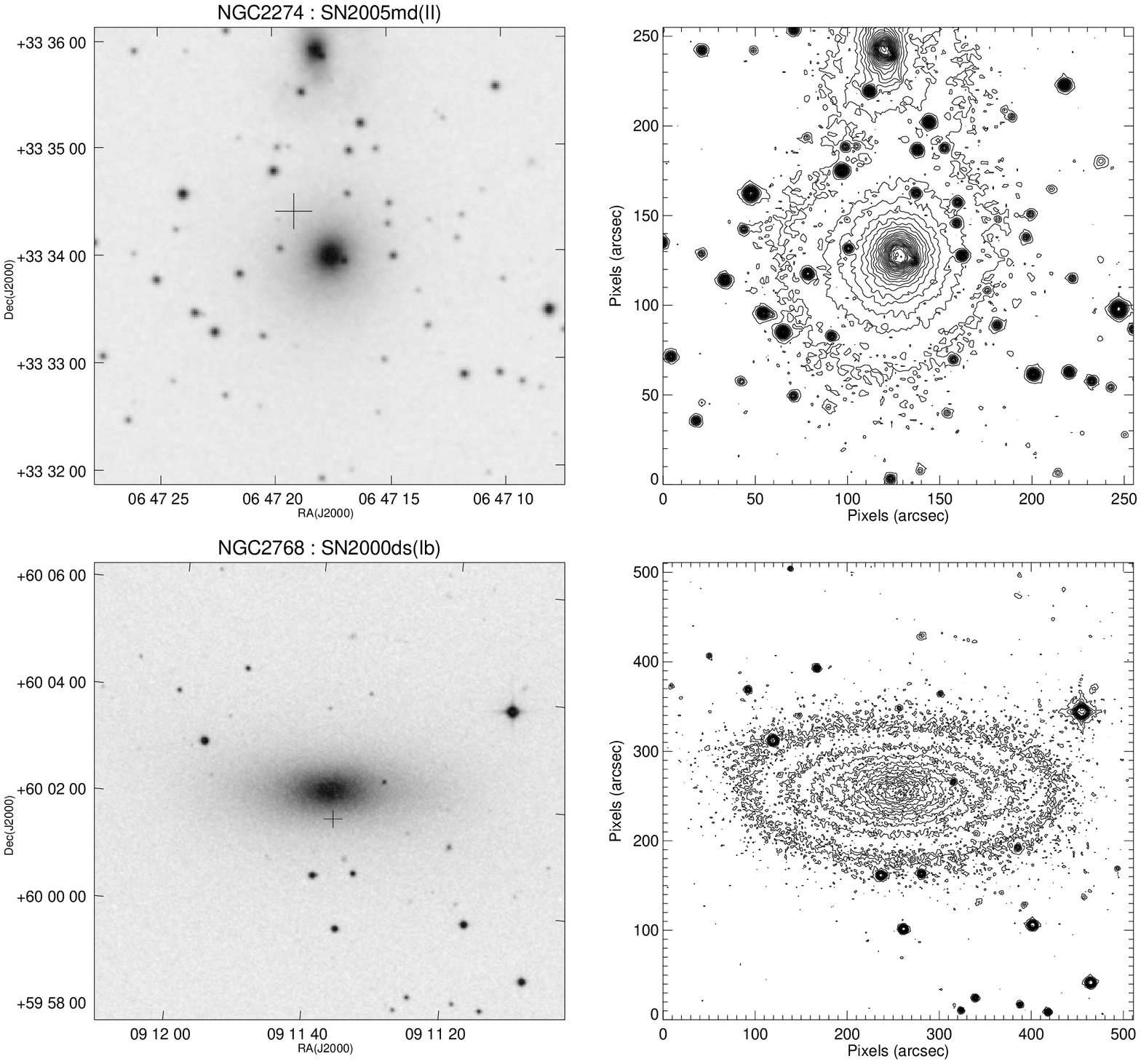}}
\caption{(c) Continued.}
\label{sn3}
\end{figure*}
\begin{figure*}[t]
\ContinuedFloat
\centering
\subfloat{
\label{sub4.1}
\includegraphics[angle=90, width=0.5\hsize]{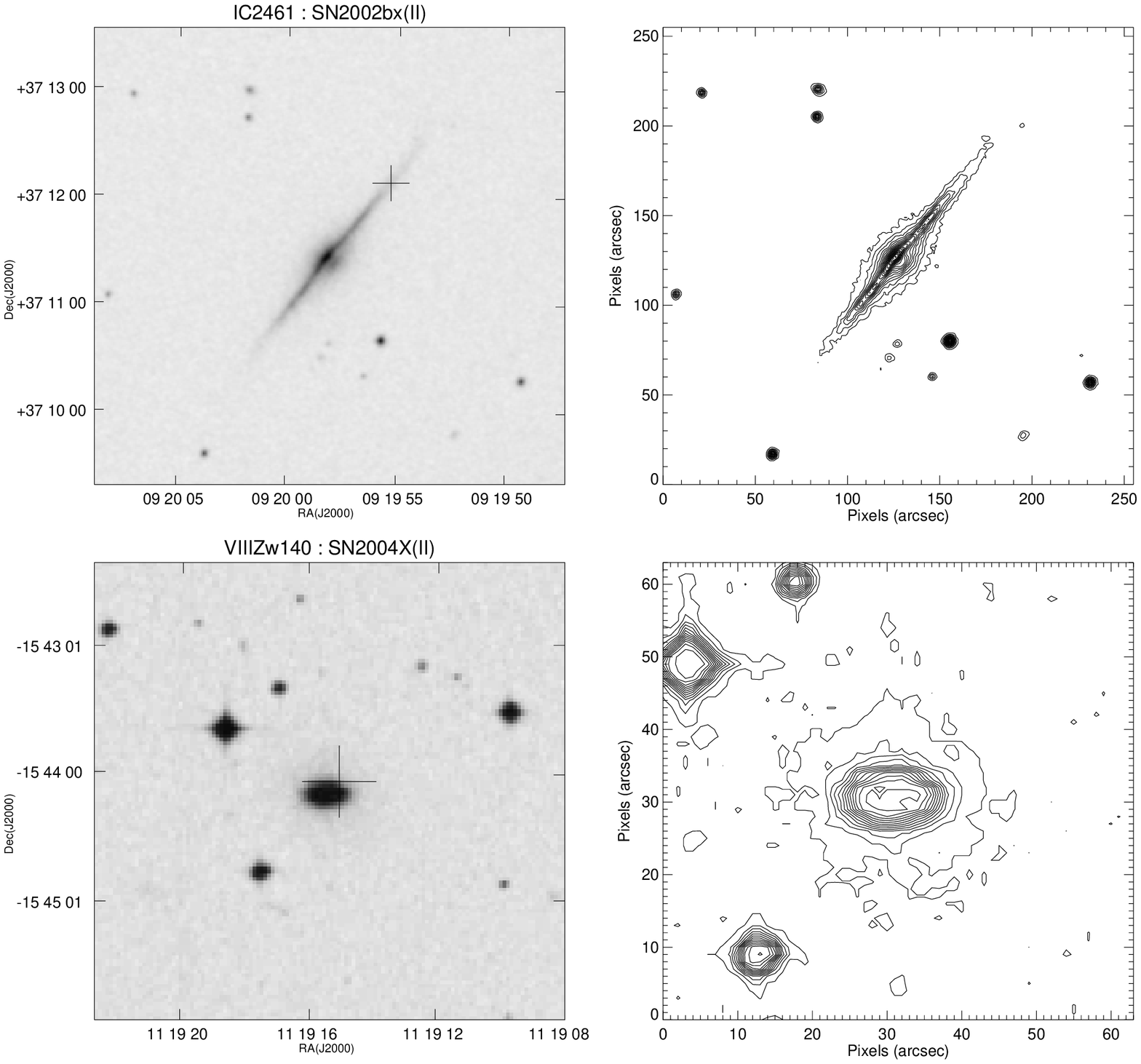}} \\
\subfloat{
\label{sub4.2}
\includegraphics[angle=90, width=0.5\hsize]{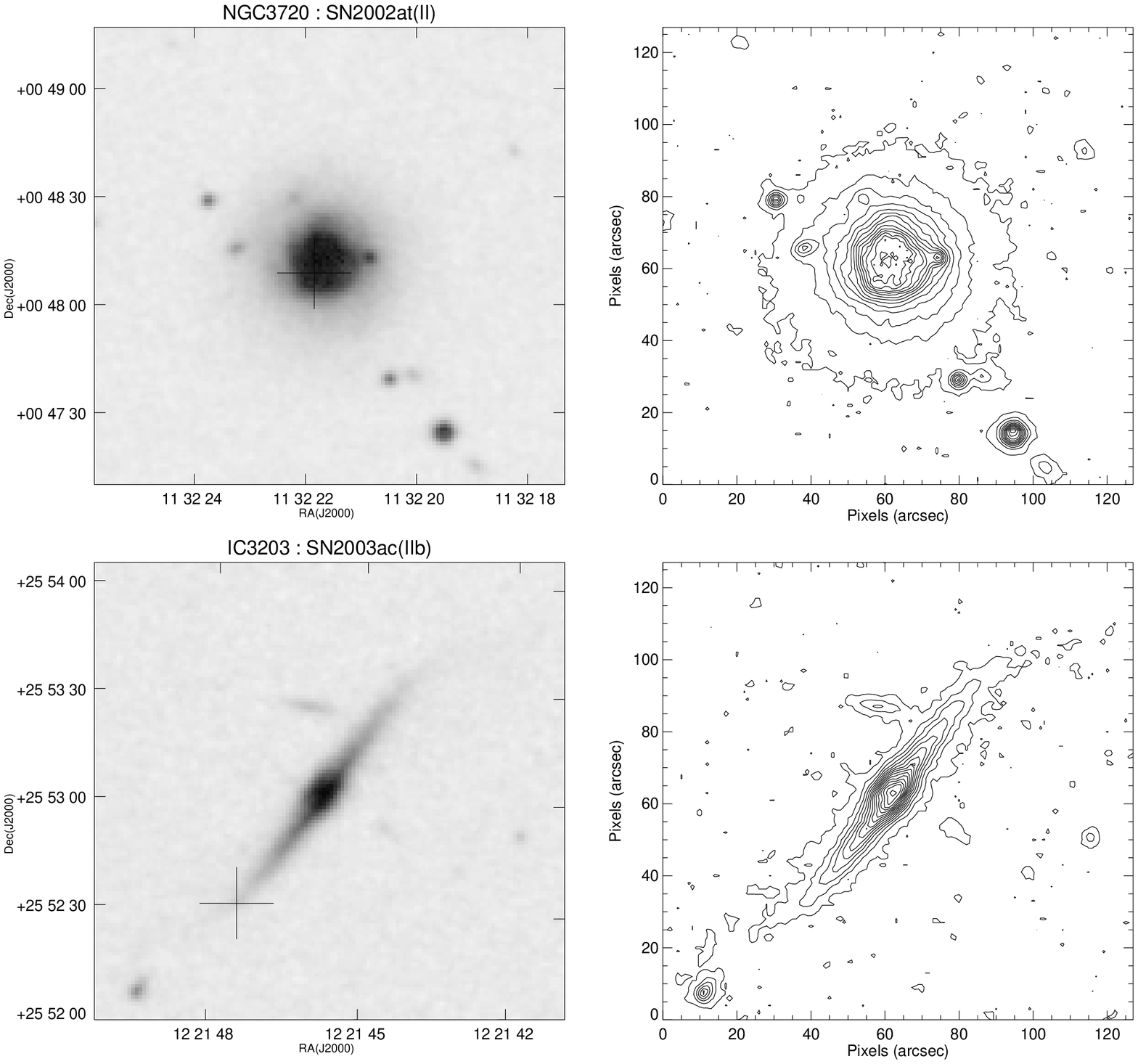}}
\caption{(d) Continued.}
\label{sn4}
\end{figure*}
\begin{figure*}[t]
\ContinuedFloat
\centering
\subfloat{
\label{sub5.1}
\includegraphics[angle=90, width=0.5\hsize]{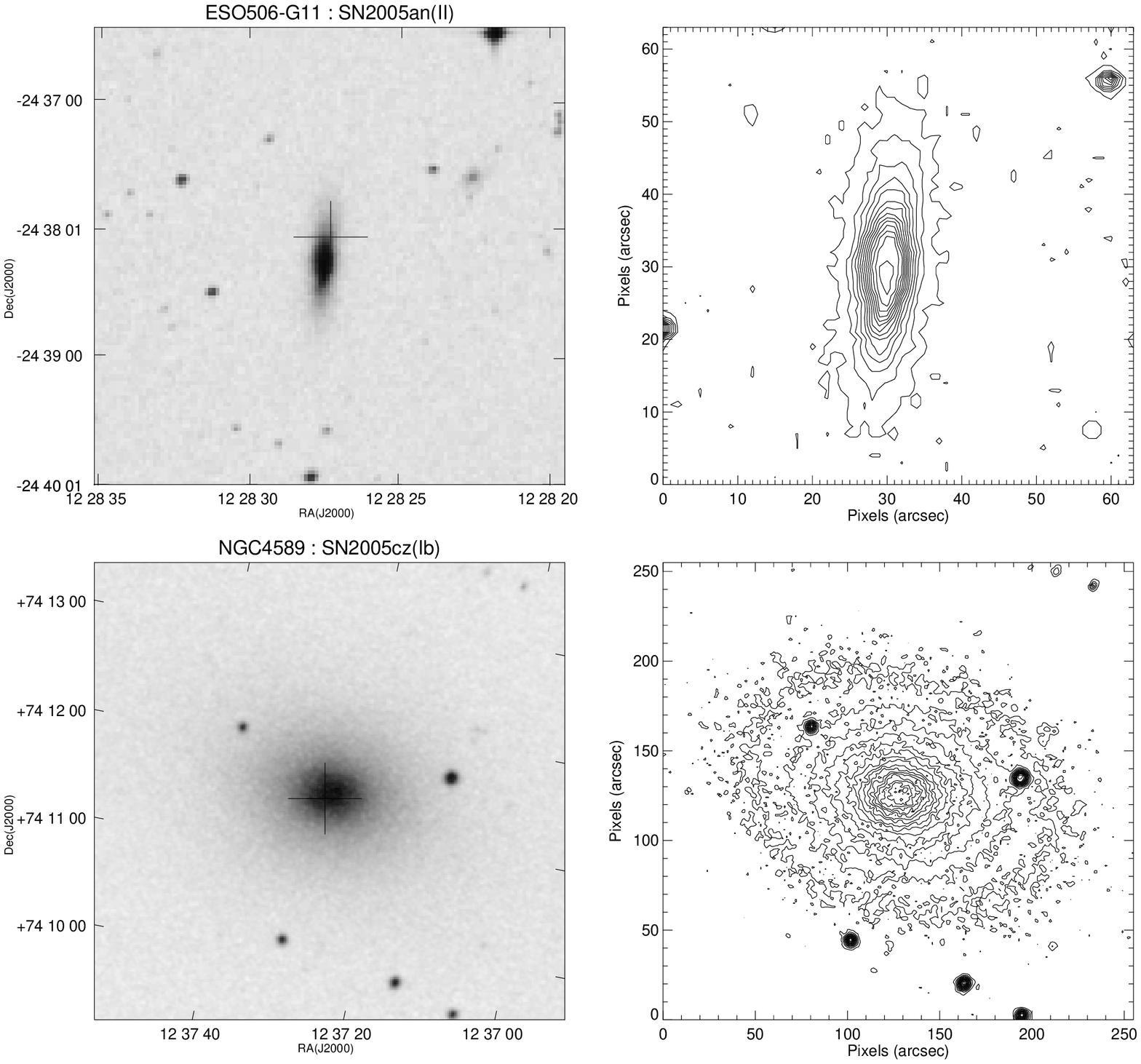}} \\
\subfloat{
\label{sub5.2}
\includegraphics[angle=90, width=0.5\hsize]{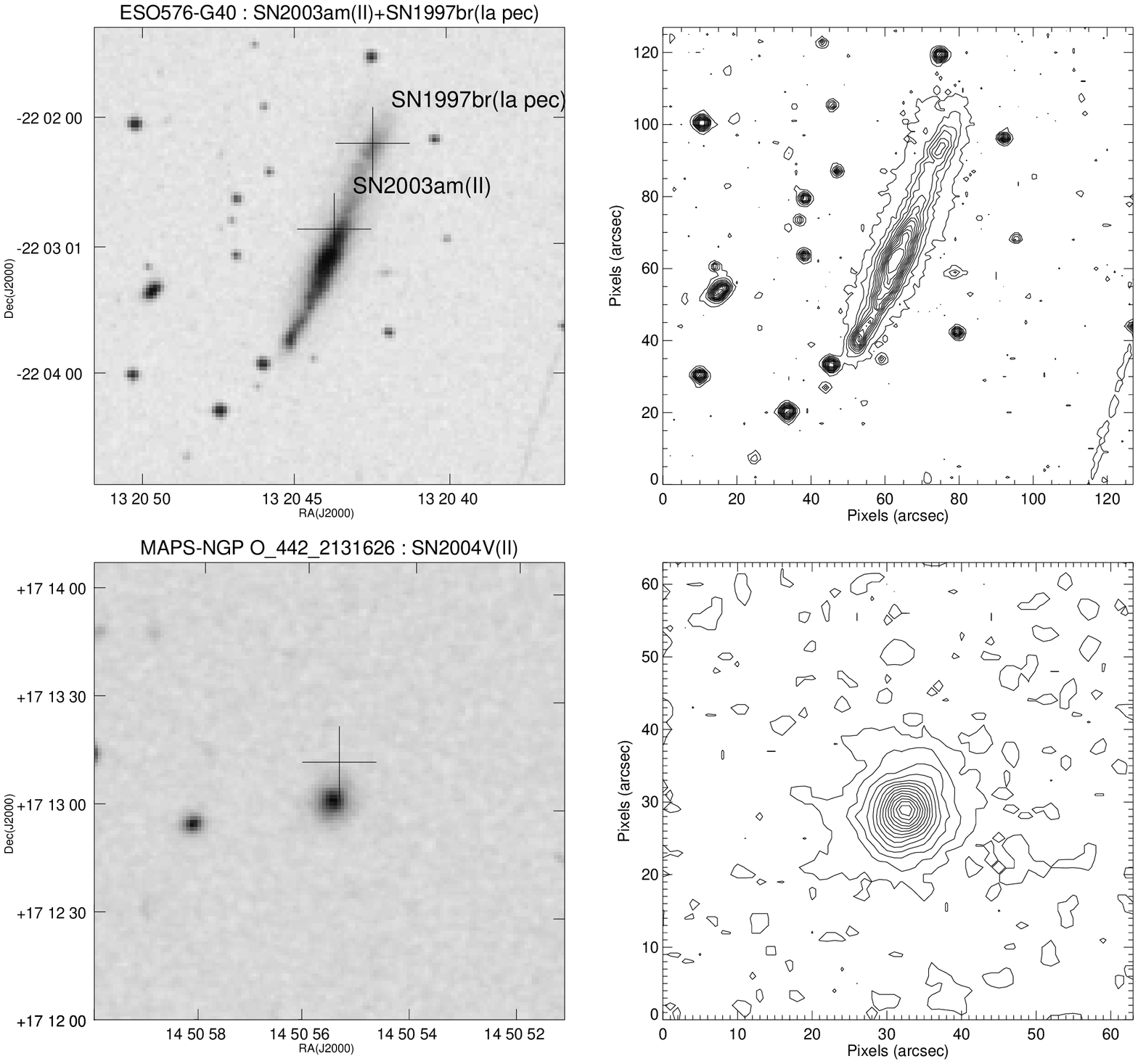}}
\caption{(e) Continued.}
\label{sn5}
\end{figure*}
\begin{figure*}[t]
\ContinuedFloat
\centering
\subfloat{
\label{sub5.1}
\includegraphics[angle=90, width=0.5\hsize]{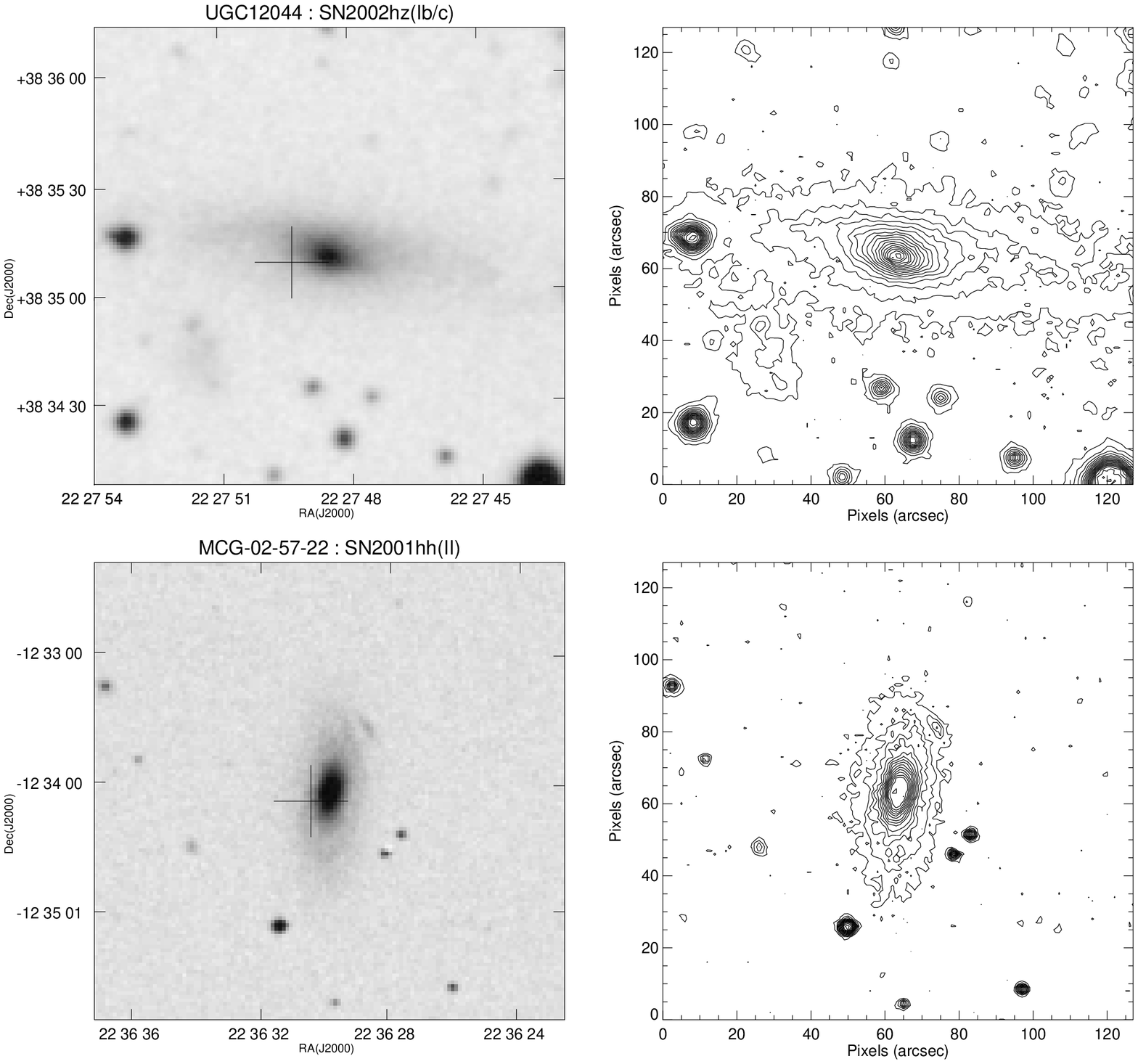}} \\
\subfloat{
\label{sub5.2}}
\caption{(f) Continued.}
\label{sn5}
\end{figure*}

\section{The morphological classification}
\label{results}

We have taken a fresh approach to classify the host galaxies in Table~\ref{tabl1}
as in \citeauthor{PMA+07} (\citeyear{PMA+07,PMA+08}). We extracted their images from the
POSS-II or SERC blue and red original plates digitally archived at STScI.
Since the original plate material was intended for the measurement of stellar
objects between 12 and 22 mag in blue and between 12 and 21 mag in red, it is
suitable for a morphological study of the sample galaxies, which are mostly
bright objects. For some very bright galaxies (brighter than 14 mag) and high
surface brightness objects, the central parts of their images may be over-exposed,
but for most galaxies the central structure is well displayed. This plate material
is therefore especially useful for examining the structural features of the galaxies
both in their inner and their outer regions. A morphological classification of each
galaxy was done first on red images and later checked on the blue ones. We have
carried out the classification by using gray-scale displays of the digitized images
and by inspecting isophotal maps. The latter ones are especially useful in representing
the large dynamical range in the images. The appearance of both the central and the
outer regions of the galaxies has been taken into account in arriving at a classification.
In Fig.~\ref{sn1} we have collected gray-scale and contour diagram representations of the \textit{F} band
images of our sample galaxies. These images are shown in figures of two images each,
comprising 22 sample galaxies. The contour levels are in arbitrary units of photographic density.
The lowest contour level was chosen at about the $3\sigma$ level of the local background.
The contour interval is constant, different in each galaxy, but usually between $10-30\%$
of the local background and chosen in order to best illustrate both the inner and the
outer structure of the galaxy. In the same fashion, the field size (and thus magnification)
was selected for each system to clearly illustrate its morphological structure. In all images,
the precise positions of the relevant SNe are also shown. These positions have been calculated
according to the offsets given in ASC and verified on the available finding charts of the SNe
when available in the WEB. For each galaxy of the sample we also conducted an extensive
literature search in order to find other information on its morphology as well as any
additional evidence for the presence of ongoing star formation. In the following we
present the individual discussion on the morphology of each host galaxy of our sample.

\textbf{IC1529.} Although listed as a S0 galaxy in the ASC, in \citet{HLdC+93}, and in
\citet{dCWP+98}, NED classifies this galaxy as  (R\textquotesingle)SA(r)0\^\,0\^\,~pec:.
This object is a UV-bright galaxy ($\rm U-B=-0.5$), with prominent $\rm H\alpha$ and [NII] emission
line spectra and it is included in the second list of the Montreal Blue Galaxy Survey
(object MBG0002-1146; \citealp{CDP+94}). According to \citet{MZW+04} this galaxy
shows strong 21~cm HI emission, and \citet{MS07} list it as a 1.4~GHz radio
continuum source. The presence of such tracers is already indicative of enhanced star
formation in this galaxy (e.g., \citealp{Ken89}). \citet{MS07} classified the
optical spectra of this galaxy as narrow LINER-like emission-line spectra. According
to our image study, we classify this galaxy as Sbc type with low surface brightness
spiral arms and with a high surface brightness, disturbed central bulge structure.
SN2006du is a classical SN~II discovered close  to maximum (\citealp{G06})
and located on an outer spiral arm.

\textbf{NGC774.} According to the ASC, this is a S0 galaxy, and the same morphological
type is reported in NED and also by \citet*{HVG99}. \citet*{CRC03}
studied the stellar population of this galaxy and concluded that there is no significant
young stellar population. This galaxy has been observed by the Sloan Digital Sky Survey
(SDSS; e.g., \citealp{A-MAA+07}), and a direct image and spectra are presented in
latest SDSS\footnote{http://www.sdss.org/dr6/} Data Release 6 (DR6) (object SDSS J015934.72+140029.5).
The nuclear region of this galaxy shows faint [NII]6584 and a trace of $\rm H\alpha$ lines in emission.
The SDSS images and our own are in agreement with previous determinations. According to its
outer isophotal structure and its radial distribution of surface brightness, this is a S0 type object.
However, the distribution of surface brightness shows some small degree of asymmetry in the region
to the south-southwest of the nucleus. Here, we suspect the presence of an embryonic spiral arm.
SN2006ee is a type II SN (\citealp{PPJ+06}) with
plateau\footnote{http://csp1.lco.cl/~cspuser1/PUB/PROJ/LC/c3\_op/SN06ee.html},
exploded very close to this region.

\textbf{NGC838.} According to ASC, NGC838 is a S0pec type galaxy and the same morphological
type (SA(rs)0\^\,0~pec) is reported in NED. The galaxy being member of Arp318 peculiar system
is also member of Hickson compact group of galaxies (HCG16C; \citealp*{HKA89}) and
classified as Im type object. According to the \citet{dCC99} it is Im type galaxy
with double nuclei structure and with narrow emission line spectra. The double nuclei structure and
irregular morphology is well seen also on the SDSS DR6 image (SDSS J020938.55-100847.5).
As an UV excess galaxy it is also known as Markarian~1022 (also is KUG0207-103) which
\citet{PMA+07} classified as Im type.
This object has blue colour ($\rm U-B=-0.06$; $\rm B-V=0.58$) and shows 1.4~GHz radio continuum
emission (\citealp{CCG+98}).
From the analysis of Fig.~\ref{sn1} we confirm double
nuclei irregular morphological structure of this galaxy. SN2005H was a normal type II discovered
short after the burst (\citealp{PTP+05}),
positioned outside the main body of this galaxy, in the low surface brightness diffuse bridge
connecting NGC838 with NGC835 (Mrk1021=HCG16A; Sab (\citealp{PMA+07}) galaxy).
In the northern edge of this bridge, closer to NGC835, an isolated HII region was
discovered (\citealp{R-WMF+04}). In all its length the bridge radiates in HI (\citealp{V-MYW+01}).

\textbf{MCG-01-07-035.} NED assigns SB(r)0+~pec type to this galaxy and LEDA classifies it as
SB0/a with outer ring (R) structure. The ASC reports it as a S0pec galaxy but
\citet*{vdBLF02} have classified it as SBab. According to
\citet{THC+07} this galaxy is a 21~cm HI source. In Fig.~\ref{sn1} this galaxy
shows an obvious bar and disturbed outer ring structure. SN2002aq, a normal SN~II
(\citealp{FB02}), was located in the inner edge of the outer ring.

\textbf{PGC10652.} According to ASC PGC10652 is an elliptical (E?), to \citet{HFL+95}
it is E/S0 while to NED it is Sb. Because of the strong HI emission \citet{HR89}
included this object in their catalogue. Our morphological study confirms the double spiral
arm structure of this galaxy and the Sb classification. SN2006ab, a type Ic SNe (\citealp{WSP+06}),
was located close to the inner edge of the northern spiral arm of the galaxy.

\textbf{IC1861.} This object is also known as the isolated galaxy KIG120 (\citealp{Ka73}).
According to ASC it is a S0 galaxy in agreement with NED (SA0\^\,0\^\,) and \citet{HVG99}
($\rm T=-2$, $\rm L-$ or $\rm S0-$). Higher resolution imaging of KIG120 led to the Sbc type classification
(\citealp{SV-MB+06}) which better agrees with its strong HI emission (\citealp{HR89}).
According to our morphological study the low surface brightness spiral structure of this galaxy
is obvious and confirms the Sbc classification. SN1999eg was a normal SN~II
(\citealp{JGCh+99}) with plateau light curve (\citealp{H01}) positioned close to the
inner edge of the spiral arm of the galaxy.

\textbf{NGC1260.} This galaxy with radial velocity of $5753\, \rm km~s^{-1}$ belongs to the Perseus
cluster (A426) and forms an interacting pair with  the close neighbor PGC12230 ($V_r=5957 \, \rm km~s^{-1}$).
ASC reports for this galaxy the S0: classification compatible with that reported by NED (S0/a:).
Early type morphology SB/SB0 ($\rm T=-1$, $\rm L+$ or $\rm S0+$) is reported also \citet{HVG99} while
according to \citet{ADP97} the morphological structure of this galaxy is similar to SA0
type objects with high amount of dust content in its inner part. An higher resolution and
sensitivity morphological study of this galaxy by \citet{MBK00} has shown weak indications
of an outer spiral and bar structure.
The presence of a short bar and a weak spiral arm structure in this galaxy is confirmed with our
imagery study. We prefer to classify this galaxy as Sa type spiral. The discovery of SN2006gy very
close to the nuclear region prompted very high resolution imaging of the parent galaxy in the optical
(Fig.~3 of \citealp{OCK+07}) and NIR (\citealp{SLF+07}). The resulting inner structure of the galaxy
is that of an early type galaxy with Sersic profile index 3.7 (\citealp{OCK+07}). Nevertheless, the
presence of a consistent dust lane and extended $\rm H\alpha$ emission (\citealp{SLF+07}) are a clear evidence
of ongoing star formation. SN2006gy is  unique in many respects and it is the brightest SN ever
observed (\citealp{OCK+07,SLF+07}; Agnoletto et al. in preparation 2008). Although
the mechanism responsible for such explosion is still debated
(\citealp{OCK+07,SLF+07,WBH07}; Agnoletto et al. 2008)
there is a general consensus that the progenitor was very massive ($M~>~40M_{\odot}$).

\textbf{IC340.} According to the ASC this is a S0pec galaxy. \citet*{vdBLF03} suspected
its spiral nature and classified it as a Spec? object. NED reports type S0\^\,0\^\, while \citet{dCWP+98}
classified it as type Sc. According to \citet{THC+07}, this object is a strong HI source.
Our inspection of the images of the galaxy confirms the spiral nature of the classification and
we assign a type Sc. The type Ic SN2002jj (\citealp{FF02}) is positioned very
close to the nuclear region.

\textbf{UGC2836.} This object is Markarian~1405, which \citet{PMA+07} classified as Sa.
The ASC considers it as a S0 and NED as a $\rm S0-$. \citet*{vdBLF02} assigned
type E1 but later decided that the classification Sa: was more appropriate (\citealp*{vdBLF03}).
According to \citet{K03} this galaxy is a strong source of neutral HI and molecular $\rm H_2$, and CO emissions
and a source of 1.4~GHz continuum radio emission (\citealp{CCG+98}).
Our morphological study confirms the Sa spiral structure of this object. Two core collapse SNe,
the type IIn SN2001I (\citealp{FCh01}) and the type Ib/c SN2003ih
(\citealp{FCh03}), have been discovered in this galaxy. SN2001I
lies closer to the nuclear region of the galaxy than SN2003ih. SN2003ih is located in the
spiral pattern of the galaxy. Identification of the background structure of the region of
the host galaxy associated with the SN2001I is problematic.

\textbf{UGC3432.} The Scd: classification reported by ASC from RC3 and by NED (\citealp{GGB+05})
and the S0/Sb classification by \citet*{vdBLF05} are contradictory enough to warrant a re-examination
of the morphology of this galaxy.
According to Huchtmeier and Richter (1989), the galaxy is a strong HI source.
This object is included in the catalog of flat disk-like galaxies by \citet{KKP93}
under the name 2MFGC5057. Our imaging study confirms the presence of late-type spiral structure
but since this object is highly inclined ($r/R=0.42$) its morphological classification
is uncertain and we prefer a Sc classification for it. Two SNe have been discovered in this galaxy,
the type Ia SN1996bv, and the type Ic: (revised classification, \citealp{FFS03}) SN2003kb.
Both SNe are projected on the disk of the galaxy.
The core collapse SN2003kb is located on a spiral arm.

\textbf{NGC2274.} The ASC and NED agree assigning type E to this galaxy.
According to \citet{Ka73} this object, together with NGC2275, is a member of
an isolated pair of galaxies (KPG118). \citet{HR89} included it in their
catalogue of HI sources. \citet{CM85} reported upper limits of $\rm 10\mu$ emission from
the nuclei of both galaxies, which is indicative of non-stellar radiation.
Sign of interaction with its northern neighbor is visible only in the outer envelope but no
sign of a tidal effect is evident. Our analysis confirms its E morphological type.
The position of SN2005md is close to the broad low-surface-brightness bridge connecting
both galaxies in the system.
The early SN classification as type II was based on the presence of a blue featureless
continuum (\citealp{MKCB+05}). About 2 years later a new SN within 0.5\arcsec from SN2005md or
a rebrightening of 2005md itself has been detected (\citealp{LFM+08}). Unfortunately no spectral
information is available to prove the nature of this object. Therefore
the possibility that the two events are related super-outbursts of a LBVit is not ruled out
and has been proposed
in the case of SN2006jc (\citealp{FSG+07,Pastorello+07}).

\textbf{NGC2768.} According to ASC this is an E6. \citet*{vdBLF03} assign type E3/Sa,
indicating doubts about its type. A dual E-Sa class was also assigned by \citet{Mau+05}.
In NED, its morphology is reported as S0\_1/2 and its activity class is LINER. The NED information
is from The Carnegie Atlas of Galaxies (\citealp{SB94}) in which the following
description is presented: ``The definite outer envelope of NGC2768 surrounding
an E6 bulge makes S0 classification certain. Very subtle dust patches on either side of the
major axis at the ends of the minor axis requires S0\_1/2 classification''.
This galaxy has blue integral colours ($\rm U-B=-0.43$ and $\rm B-V=0.92$) and
it is registered as a CO (\citealp{WCH95}) and HI emission source (\citealp{HSH95})
and a source of 1.4~GHz radio continuum emission (\citealp{CCG+98}).
Perhaps the most interesting information relates to
the nuclear and near-nuclear regions of this galaxy: according to \citet{MM94}
``\ldots the inner isophotes of the galaxy are boxy, but become disky in the range of $75-115$ arcsecs''.
According to \citet{LFG+05} and \citet{MES+06} this galaxy has a dust-obscured
central region (spiral dust lanes), central disky gas distribution and a LINER nucleus with strong
emission lines. The stellar population in the central region of this galaxy is relatively young.
The resolution of our images was not sufficient to analyze in detail the central nuclear structure, but
from its outer structure this object is a classical S0 type.
SN2000ds, classified as type Ib (\citealp{FCh00}) is located
far from the peculiar nuclear region of the galaxy. Detailed study of the SN position by
\citet*{vD+03} shows that ``\ldots reddish clusters of stars may be evident
near the position of the SN at the $3-4~\sigma$ level''.

\textbf{IC2461.} \citet*{vdBLF03} classified this galaxy as S0,
but noted that it is viewed edge-on and so the classification is somewhat uncertain.
ASC classified it as Sb, and NED assigned Sbc. This galaxy is included in the catalog of
flat disk-like galaxies (\citealp{KKP93}) under the name of 2MFGC7243 and, according
to \citet{GR81}, has a well-defined regular velocity field typical for late-type
spiral galaxies. According to images and spectra from the SDSS DR6 (SDSS J091958.02+371128.5)
it is an edge-on Sbc galaxy with well-delineated dust lanes and with spectra rich in strong
emission lines ($\rm H\alpha$, [OIII], etc.). This object is a strong source of HI emission
(\citealp{THC+07}) and 1.4~GHz radio continuum emissions (\citealp{CCG+98}).
Our imaging study confirms the Sbc morphological classification.
SN2002bx is a type II SN discovered soon after maximum
(\citealp{MJCh+02}) and positioned in the far edge of the disk.

\textbf{VIIIZw140.} No galaxy type is reported in the ASC for this galaxy which was classified
as E3 by \citet*{vdBLF05}. NED assigned type E.
According to \citet{PTB+03}, it is an HI galaxy
and also has 1.4~GHz continuum radio emission (\citealp{CCG+98}).
In our images, the central region of
the galaxy are over-exposed whereas in the outer regions the surface brightness declines from
a regular distribution and forms elongations to the north and south and the existence
of two embryonic spiral arms is suspected. SN2004X, classified as a SN~II
(\citealp{FChF04}), is positioned in the outer envelope of the galaxy.
No specific morphological structure is present at the SN position.

\textbf{NGC3720.} \citet*{vdBLF03} classified this galaxy as E1, but noted that its type
might be Sa, which is the ASC classification. This galaxy is listed as SAa in NED,
and \citet{K73} identified it as a member of an isolated pair (KPG289B).
According to \citet{SB94} ``The nucleus of this galaxy is small and bright.
The multiple arms are smooth at the resolution of the present plate material.
The galaxy is clearly a spiral''. \citet*{ShDP06} described it as SAa with emission line spectra.
The blue integral colours ($\rm U-B=-0.04$ and $\rm B-V=0.76$) of the galaxy are typical
for Sb spirals (e.g., \citealp{H1977}).
According to \citet{PTB+03} this galaxy has detectable HI emission
and according to \citet{CCG+98} is 1.4~GHz continuum radio emission source.
This object has
been observed in the SDSS (SDSS J113221.83+004816.5) and it is obviously an Sa-type spiral galaxy.
In our images the nuclear regions are over-exposed, but the structure of the
circum-nuclear and outer regions confirms its spiral nature. We suggest a Sa classification.
SN2002at, classified as possible young SN~II  (\citealp{FB02}),
appeared on an outer narrow spiral arm of the galaxy.

\textbf{IC3203.} \citet*{vdBLF03} classified this galaxy as S0: but noted, as for IC2461,
that it is viewed edge-on and therefore has a somewhat uncertain classification. In both
the ASC and NED it is reported as Sb. In the Gold Mine Database (e.g., \citealp{GBC+06})
it is described as an Sb galaxy with emission line spectra. This galaxy is included in the
catalog of 2MASS-selected flat disk-like galaxies (\citealp{MKK+04}) under the name of
2MFGC9726. According to \citet{PTB+03} it is an HI object. Our imaging study confirms
the Sb morphological type for this galaxy. SN2003ac, confirmed to be a rare  SN~IIb
(\citealp{FiCh03}), is located in the outer parts of the disk.

\textbf{ESO506-G011.} ASC and NED both report this as an S0 galaxy. \citet{dCWP+98} agree,
assigning it ($-2$) S0 type. The analysis of this edge-on object shows the presence of a prominent
bulge and disk structure. We suggest to classify it as an early (possibly Sb type) spiral galaxy.
SN2005an was an early SN~II (\citealp{MKCh+05}) located in the disk of this galaxy.

\textbf{NGC4589.} According to the ASC and NED this is an E2 galaxy. Our images do show
a surface brightness distribution resembling an elliptical galaxy. According to \citet{THC+07}
it is a strong 21~cm HI source. In a study of the gas and stellar kinematics of this galaxy,
\citet{RHP+01} concluded that it is a merger remnant (see also \citealp{MB89})
with LINER-type nuclear activity (\citealp{LAK+98}).
Studies by \citet{MM94}, \citet{CFI+97} and \citet{LFG+05}
show the existence of large, chaotic, and disorganized opaque dust patches
(in and outside of 4\arcsec fields), forming asymmetric patterns, which might indicate a disk.
Summarizing above data we conclude that this galaxy is a merger remnant with an outer structure
similar to an elliptical galaxy.
The object is a source of 1.4~GHz continuum radio emission (\citealp{CCG+98}).
SN2005cz is a type Ib SN (\citealp{L05}) 17\arcsec far from
the nucleus. The peculiarities in the central regions of this galaxy
(and reaching out to the location of the SN) are possibly associated with star
formation (e.g., \citealp{KHT+03}).

\textbf{ESO576-G040.}  According to \citet*{vdBLF03} this is an S0t type galaxy.
NED assigned type SBd?~pec. According to \citet*{CdVdV85} ``Galaxy has over-exposed bar?
Pretty smooth north arm, very knotty south arm''. \citet{GGB+05} classified this object
as an SBd type spiral with emission line spectra. Following \citet{MZW+04} it is HIPASS HI source J1320-22.
Our imaging of this highly-inclined galaxy confirms its spiral nature. A bar and two spiral arm
structures can be identified therefore we prefer to classify this galaxy as SBc type. In this galaxy
two SNe have been discovered: the type II SN2003am (\citealp{FiCh03})
located on a bright condensation of a northern spiral arm and the peculiar type Ia SN1997br
(\citealp{LQQ+99}) in the disk of the galaxy to the north of the nucleus
and close to another bright HII region.

\textbf{MAPS-NGP O\_442\_2131626.} \citet*{vdBLF05} classified this galaxy as E0,
the same morphological type (E:0) assigned by NED. In the literature there are no published
data. Our morphological study reveals a low-contrast spiral arm structure
pointing in a north/north-west direction. We classify the galaxy as Sa type. The type II SN2004V
(\citealp{FChF04}) is positioned close to the faint spiral arm structure.

\textbf{UGC12044.} \citet*{vdBLF03} classified this galaxy as Sb/S0.
The ASC, NED, and \citet*{HVG99} all agree that this is an Sab galaxy.
According to \citet{PTB+03} HI has been detected in it. Our imaging study confirms
the Sab morphological classification. The type Ib
SN2002hz (\citealp{FiCho03}) is projected on the disk
but not associated with particular structures.

\textbf{MCG-02-57-22.} In the ASC this galaxy is classified as S0, whereas according
to \citet*{vdBLF05} it is an Sa. NED assigned an
(R\textquotesingle)SB0\^\,0\^\,~pec? morphological type while according to \citet{KaKa00}
it is an Sa, and a member of an isolated triplet (KTS70A). \citet{dCWP+98} classified
it as Im~(10) galaxy. Following to \citet{FrKo89} it is a faint $\rm H\alpha$ emitter.
Our morphological study confirms the spiral nature of this galaxy (Sa classification).
The Las Campanas 6.5~m telescope
image\footnote{http://cfa-www.harvard.edu/supernova/images/2001hh.jpg}
of SN2001hh, a normal type II SN (\citealp{MJK+01}),
shows it on the inner edge of a southern spiral arm, thus supporting our claim.

\section{Discussion and Conclusions}
\label{discus}

It is generally believed (e.g., \citealp{vdBT+91,CET99})
that the hosts of core collapse SNe are objects with young stellar population
(generally spiral or irregular galaxies) while
the old stellar population of early type galaxies can produce only SNe~Ia.
Nevertheless among the morphologically-classified host galaxies
of CCSNe, we have noted 22 cases in which the host has been classified as elliptical or S0,
in apparent contradiction to this conventional view. Indeed it is well known that the
morphological classification
of galaxies has margins of subjectivity. For instance \citet*{vdBLF03} have noted
that ``\ldots late type galaxies of unusually high surface brightness may be misclassified as being
of early type'', and mention several additional minor issues with the classifications.

For the 22 cases in which the host
galaxies of CCSNe have been reported as early types we have assembled imaging and searched the
literature for additional information. We found that 14 hosts clearly ought to be re-classified as spirals,
for three galaxies the presence of spiral structure is suspected, one galaxy has been classified as irregular,
one as a barred galaxy with an outer ring structure.
We share the view with \citet*{vdBLF03} that  high surface brightness
in  galaxies plays a crucial role in misclassification, particularly for largely inclined
galaxies. There are no detailed studies on the surface brightness of our sample galaxies
but we outline that six out of 22 galaxies are largely inclined objects.
Since radial velocity may be another important factor influencing
morphological classification
we have compared the mean radial velocities
of the total sample of CCSNe hosts in ASC with available recession velocity and morphology
(910 galaxies) with those of the sample of 22 galaxies. The mean radial velocity of the total sample
is $4935\pm3153\, \rm km~s^{-1}$ to be compared with $4996\pm2434\, \rm km~s^{-1}$ of our galaxies.
This means that the uncertain classification was not due to the distance effect.

Other parameters, in addition to the morphological type, bar and ring are indicators of young stellar
populations. In the sample there is one merger (NGC4589) with elliptical outer isophotes and
active nuclear region, which may also have enhanced star formation
(e.g., \citealp{KHT+03}). In fact, the SN in this galaxy is close to the nucleus.
Another galaxy (NGC2768) has been confirmed as S0 with a central region showing dust and a disky
central gas distribution.
This galaxy has a powerful LINER nucleus, and may also be the remnant of a merger.
The existence of strong HI and CO emission, and the presence of dust and ionized gas supports the
presence of a young stellar population. The core collapse SN in this galaxy is possibly
associated with the red star clusters present. Only one galaxy of the 22, NGC2274, remains confirmed
with an elliptical morphology without any inner structural peculiarity. This galaxy is a member of
an isolated pair, and appears to be in interaction with a close neighbor spiral.
NGC2274 may possibly have accreted some gas during a tidal encounter with this gas-rich neighbor,
thereby accounting for the presence of massive stars.
Despite the low number statistics it is interesting to note the unusually high ratio
$\rm SN~Ibc~/~SN~II=2/1$ found in early type galaxies with respect to other  galaxy types
(e.g., Table~5 of \citealp{BBCT99}).

Our sample of 22 galaxies includes 17 HI sources (more than 77\% of the total sample).
It has been shown many times that the presence of HI emission is correlated with star formation,
and the HI is generally found in close proximity to star-forming regions (e.g., \citealp{Hel07}).
A high HI detection rate among our sample galaxies is consistent with the presence of a young stellar
population in these objects.
Our sample includes also 8 1.4~GHz continuum radio sources. According to the criterion
adopted by \citet{DVPP+05},
among  8 radio sources detected 2 (NGC838, UGC2836) are radio-loud and 6 (IC1429, NGC2768,
NGC3720, IC2461, VIIIZw140 and NGC4589) are radio-faint. In two cases (NGC2768, NGC4589) the
LINER AGNs might contribute to the radio emission (e.g., \citealp{FNW+00}).
In the 6 remaining cases the radio emission might witness recent star
formation (e.g., \citealp{R-GST+07}).
It is interesting to note that two galaxies in the sample
are UV excess Markarian galaxies, and three have LINER type AGNs. In one case (NGC1260 with SN2006gy),
the SN is located in the inner core of the galaxy characterized by extended $\rm H\alpha$ emission,
consistent with the supposed large mass of the progenitor.

In summary:

\begin{enumerate}
  \item In 17 cases, the ``early type'' host is a misclassified spiral galaxy.
  \item In one case the host galaxy is a misclassified irregular, and in another
        it is a misclassified ring galaxy.
  \item Three galaxies retain the early type classification: NGC4579 a merger remnant with
        elliptical outer isophotes; NGC2768, a S0 with LINER nucleus, possibly the remnant of a merger;
        NGC2274 an elliptical with no morphological peculiarity which anyway appears to be in close
        interaction with a neighbour.
\end{enumerate}

In conclusion, the presence of CCSNe in early type galaxies can be interpreted as an additional
indication that residual star formation episodes take place also in E and S0 galaxies due
to merging/accretion or interaction with close neighbours. In other words the morphology of a galaxy
is the most direct and immediate indicator of its general stellar content and star formation history.
Nevertheless the morphology is inadequate to describe local SF episodes occurring at small scale.

The available small statistics is inadequate to perform a detailed analysis of the frequency
of CCSNe in early type galaxies, hence to test whether the presence of CCSNe is compatible with
the existence of a \textit{``prompt''} population of SN~Ia as invoked by \citet{MDVP06}.

\begin{acknowledgements}

A.H. acknowledges the hospitality of the Institut d'Astrophysique de Paris (France).
The work of A.~A.~Hakobyan at IAP was supported by
PICS France-Armenie and the grant of French Government. A.P. acknowledgments the hospitality
of the Space Telescope Science Institute during his stay as visiting scientist supported by
the Director's Discretionary Research Fund.
A.P. wishes to thank the Institut d'Astrophysique de Paris (France)
hospitality and support (via the EARA network) during the last stage of this work.
M.T. acknowledge the support of the Italian
Ministry of University and Research via PRIN RIN 2006 n.2006022731~002. Funding for the SDSS
and SDSS-II has been provided by the Alfred P.~Sloan Foundation, the Participating Institutions
are the National Science Foundation, the U.S. Department of Energy, the National Aeronautics and
Space Administration, the Japanese Monbukagakusho, the Max Planck Society, and the Higher Education
Funding Council for England. The SDSS Web Site is http://www.sdss.org/. This research has made use
of NASA/IPAC Extragalactic Database (NED), which is operated by the Jet Propulsion Laboratory,
California Institute of Technology, under contract with the National Aeronautics and Space Administration,
and HyperLeda (Leon-Meudon Extragalactic Database, http://cismbdm.univ-lyon1.fr/$\sim$hyperleda/).
The Digitized Sky Survey was produced at the Space Telescope Science Institute under U.S.
Government grant NAG W--2166. The images of this survey are based on photographic data
obtained using the Oschin Schmidt Telescope on Palomar Mountain and the UK Schmidt Telescope.
The plates were processed into the present digital form with the permission of these institutions.
Finally, we are especially grateful to our referee for promptly and carefully reading the
manuscript and making several constructive comments which have improved the paper quite significantly.

\end{acknowledgements}

\end{document}